\documentclass[12pt]{iopart}
\usepackage{graphicx}
\usepackage{amssymb}
\let\Newcommand=\newcommand
\def\newcommand{\providecommand}
\let\newcommand=\Newcommand

\usepackage{color}
\usepackage{fancyheadings}
\usepackage[pdftitle={Gravitational solution to the Pioneer 10/11 anomaly}, pdfauthor={J. R.
Brownstein and J. W. Moffat},
bookmarks, bookmarksopen, colorlinks=true, linkcolor=blue, citecolor=blue, anchorcolor=blue, urlcolor=blue]{hyperref}
\newcommand{\au}{\,\mbox{AU}}
\newcommand{\preprint}[1]{\href{http://arxiv.org/abs/#1}{{\it Preprint} #1}}
\newcommand{\citeasnoun}[1]{Ref.\,\cite{#1}}
\newcommand{\JCAP}{{\it J. Cosmol. Astropart. Phys.}\ }
\newcommand{\MNRAS}{{\it Mon. Not. R. Astron. Soc.}\ }
\newcommand{\ApJ}{{\it Astroph. J.}\ }
\newcommand{\SSR}{{\it Solar Sys. R.}\ }

\begin{document}
\title{Gravitational solution to the Pioneer 10/11 anomaly}
\author{J.\,R. Brownstein and J.\,W. Moffat}
\address{Perimeter Institute for Theoretical Physics, Waterloo, Ontario, N2L 2Y5, Canada}
\address{Department of Physics, University of Waterloo, Waterloo, Ontario N2L 3G1, Canada}
\ead{\href{mailto:jbrownstein@perimeterinstitute.ca}{jbrownstein@perimeterinstitute.ca}, \href{mailto:john.moffat@utoronto.ca}{john.moffat@utoronto.ca}}

\begin{abstract}
A fully relativistic modified gravitational theory including a fifth force skew symmetric field is fitted to the Pioneer
10/11 anomalous acceleration. The theory allows for a variation with distance scales of the gravitational constant $G$,
the fifth force skew symmetric field coupling strength $\omega$ and the mass of the skew symmetric field
$\mu=1/\lambda$. A fit to the available anomalous acceleration data for the Pioneer 10/11 spacecraft is obtained for a
phenomenological representation of the ``running'' constants and values of the associated parameters are shown to exist
that are consistent with fifth force experimental bounds. The fit to the acceleration data is consistent with all
current satellite, laser ranging and observations for the inner planets. 
\end{abstract}
\published{\href{http://www.iop.org/EJ/abstract/0264-9381/23/10/013}{Class. Quantum Grav. {\bf 23} (2006) 3427-3436}}
\section{Introduction}
The radio tracking data from the Pioneer 10/11 spacecraft 
during their travel to the outer parts of the solar system have revealed an anomalous acceleration. The Doppler data
obtained at distances $r$ from the Sun between $20$ and $70$ astronomical units (AU) showed the anomaly as a deviation
from Newton's and Einstein's gravitational theories. The anomaly is observed in the Doppler residuals data, as the
differences of the observed Doppler velocity from the modelled Doppler velocity, and can be represented as an anomalous
acceleration directed towards the Sun, with an approximately constant amplitude over the range of distance, $20\au < r <
70\au $~\cite{Anderson,Anderson2,Turyshev}:
\begin{equation} \label{aP}
a_P=(8.74\pm 1.33)\times 10^{-8}\,{\rm cm}\,s^{-2}.
\end{equation}
After a determined attempt to account for all {\it known} sources of systematic errors, the conclusion has been reached
that the anomalous acceleration towards the Sun could be a real physical effect that requires a physical
explanation~\cite{Anderson,Anderson2,Turyshev}.

Two theories of gravity called the metric-skew-tensor gravity (MSTG) theory~\cite{Moffat} and the scalar-tensor vector
gravity (STVG) theory~\cite{Moffat2}  have been proposed to explain the rotational velocity curves of galaxies, clusters
of galaxies and cosmology. A fitting routine for galaxy rotation curves has been used to fit a large
number of galaxy rotational velocity curve data, including low surface brightness (LSB), high surface brightness (HSB)
and dwarf galaxies with both photometric data and a two-parameter core model without non-baryonic dark
matter~\cite{Moffat2,Brownstein}. The fits to the data are remarkably good and for the photometric data only the one
parameter, the stellar mass-to-light ratio, $\langle M/L\rangle$, is used for the fitting, once two parameters $M_0$ and $r_0$
are universally fixed for galaxies and dwarf galaxies. A large sample of X-ray mass profile cluster data has also been
fitted~\cite{Brownstein2}.

The gravity theories require that Newton's constant G, the coupling constant $\omega$ that measures the strength of the
coupling of the skew field to matter and the mass $\mu$ of the skew field, vary with distance and time, so that
agreement with the solar system and the binary pulsar PSR 1913+16 data can be achieved, as well as fits to galaxy
rotation curve data and galaxy cluster data. In \citeasnoun{Moffat2} and \citeasnoun{Reuter}, the variation of these
constants were based
on a renormalization group (RG) flow description of quantum gravity theory formulated in terms of an effective classical
action. Large infrared renormalization effects can cause the effective $G$, $\omega$, $\mu$ and the cosmological
constant $\Lambda$ to run with momentum $k$ and a cutoff procedure leads to a space and time varying $G$, $\omega$ and
$\mu$, where $\mu=1/r_0$ and $r_0$ is the effective range of the skew symmetric field. In the STVG
theory~\cite{Moffat2}, the action contains contributions that lead to {\it effective} field equations that describe
the variations of $G$, $\mu$ and the coupling constant $\omega$ that measures the strength of the coupling of the skew
symmetric field $B_{\mu\nu}$ with matter. In principle, we can solve for the complete set of field equations and
determine the dynamical behavior of all the fields. However, in practice we make approximations allowing us to obtain
partial solutions to the equations, yielding predictions for the various physical systems considered.

Both the MSTG and STVG theories lead to the same modified acceleration law for weak
gravitational fields and the same fits to galaxy rotation curve and galaxy cluster data, as well as to agreement with
the solar system and pulsar PSR 1913+16 observations.  However, the STVG theory is simpler in its structure, so in the
following, we shall restrict our attention to this theory.

An important constraint on gravity theories is the bounds obtained from weak
equivalence principle tests and the existence of a ``fifth'' force, due to the exchange of a massive vector
boson~\cite{Fischbach,Talmadge,Adelberger,Adelberger1,Adelberger2,Adelberger3}. These bounds cannot rule out gravity
theories that violate the weak
equivalence principle or contain a fifth force at galactic and cosmological distance scales. However, we shall in the
following attempt to explain the anomalous acceleration observed in the Pioneer 10/11 spacecraft
data~\cite{Anderson,Anderson2,Turyshev} and simultaneously fit the solar planetary data. For our study of the solar
system, we must account in our modified gravity theory not only for the variation of $G$ with respect to the radial
distance $r$ from the center of the Sun, but also for the variations of $\mu=1/r_0$ and $\omega$ with respect to $r$.
Since we do not possess rigorous solutions for the variations of $r_0$ and $\omega$, we use a phenomenological
parameterization of the varying parameters $r_0$ and $\omega$ to obtain fits to the anomalous Pioneer acceleration data
that are consistent with the solar system and fifth force experimental bounds. As has been demonstrated by  \citeasnoun{Reynaud}, the fifth force experimental bounds rule out a gravitational and fifth force explanation of the
Pioneer anomaly for fixed universal values of the coupling strength $\alpha$ and the range $\lambda=r_0$.

In an asymptotically free RG flow description of quantum gravity, a distance (momentum) scaling law for the
gravitational coupling constant $G$, the coupling constant $\omega$ and the range $\lambda$ leads to strong infrared
renormalization effects for increasing distance scale~\cite{Reuter,Moffat}. We shall find in the following that an
asymptotically free-type of gravity and fifth force distance scaling behavior can possibly explain the anomalous Pioneer
acceleration and be in agreement with local E\"otv\"os experiments and planetary deep space probes.

An important feature of the MSTG and STVG theories is that the modified acceleration law for weak gravitational fields
has a {\it repulsive} Yukawa force added to the Newtonian acceleration law for equal ``charges''. This corresponds to
the exchange of a massive spin-1 boson, whose effective mass and coupling to matter can vary with distance scale. A
scalar component added to the Newtonian force law for equal charges would correspond to an {\it attractive} Yukawa force
and the exchange of a spin-0 particle. The latter acceleration law cannot lead to a satisfactory fit to galaxy rotation
curves and galaxy cluster data~\cite{Brownstein,Brownstein2}.

All the current applications of the two gravity theories that can be directly confronted with experiment are based on
weak gravitational fields. To distinguish the MSTG and STVG theories, it will be necessary to obtain experimental data
for strong gravitational fields, for example, black holes as well as cosmology.

\section{Weak Fields and the Modified Gravitational Acceleration}

The equation of motion of a test particle, about a mass $M$, obtained for weak gravitational fields and for a static
spherically symmetric solution of the massive skew field $B_{\mu\nu}$ in the STVG theory takes the form~\cite{Moffat2}:
\begin{equation}
\label{particlemotion2}
\frac{d^2r}{dt^2}-\frac{J^2_N}{r^3}+\frac{GM}{r^2}=K\frac{\exp(-\mu r)}{r^2}(1+\mu r),
\end{equation}
where $K$ and $\mu$ are positive constants and $J_N$ is the Newtonian angular momentum. We have not assumed in 
\Eref{particlemotion2} any particular composition dependent model for $K$ for the additional Yukawa force term.

We observe that the additional Yukawa force term in \Eref{particlemotion2} is {\it repulsive} for like charges
in
accordance with the exchange of a spin-$1^-$ massive boson. We shall treat $\alpha=\alpha(r)$ and $\mu=1/\lambda(r)$
as running parameters which are {\it effective} functions of $r$ and obtain for the radial acceleration on a test
particle
\begin{equation}
\label{acceleration} a(r)=-\frac{G_{\infty}M}{r^2}+K(r)\frac{\exp(-r/\lambda(r))}{r^2}
\biggl(1+\frac{r}{\lambda(r)}\biggr).
\end{equation}
Here, we are treating the ``constants'' $K$ and $\lambda$ as running with distance scale, in accord with our effective RG
flow interpretation of $G(r)$, $K(r)$ and $\lambda(r)$~\cite{Moffat,Reuter}. On the other hand, in the STVG
theory~\cite{Moffat2} the varying constants are treated as scalar fields and are  solved in terms of effective field
equations obtained from an action. In \Eref{acceleration}, $G_{\infty}$ denotes the renormalized value of the
gravitational constant:
\begin{equation}
\label{renormG} G_{\infty}=G_0\biggl(1+{\alpha_{\infty}}\biggr),
\end{equation}
where $G_0 = 6.674 \times 10^{-8} \mbox{g}^{-1}\mbox{cm}^{3} \mbox{s}^{-2}$ is Newton's ``bare'' gravitational constant.
Let
\begin{equation}
\label{Kequation} K(r)=G_0M\alpha(r).
\end{equation}

By using \Eref{renormG}, we can rewrite the acceleration in the form
\begin{equation}
\label{accelerationlaw} a(r)=-\frac{G_0 M}{r^2}\biggl\{1+\alpha(r)\biggl[1-\exp(-r/\lambda(r))
\biggl(1+\frac{r}{\lambda(r)}\biggr)\biggr]\biggr\}.
\end{equation}

The acceleration law (\ref{accelerationlaw}) can be written
\begin{equation}
\label{accelerationGrun} a(r)=-\frac{G(r)M}{r^2},
\end{equation}
where
\begin{equation}
\label{runningNewton} G(r)=G_0\biggl[1+\alpha(r)\biggl(1-\exp(-r/\lambda(r))
\biggl(1+\frac{r}{\lambda(r)}\biggr)\biggr)\biggr]
\end{equation}
is an {\it effective} expression for the variation of $G$ with respect to $r$.

\section{Pioneer Anomalous Acceleration}

We postulate a gravitational solution that the Pioneer 10/11 anomaly is caused by the difference between
the running $G(r)$ of \Eref{runningNewton} and the bare value, $G_{0} = 6.674 \times 10^{-8} \mbox{g}^{-1}\mbox{cm}^{3} \mbox{s}^{-2}$.  So the
Pioneer anomalous acceleration directed towards the center of the Sun is given by
\begin{equation}
a_P=-\frac{\delta G(r)M_{\odot}}{r^2},
\end{equation}
where
\begin{equation} \label{deltaG}
\delta G(r)=G_0\alpha(r)\biggl[1-\exp(-r/\lambda(r))
\biggl(1+\frac{r}{\lambda(r)}\biggr)\biggr].
\end{equation}
We propose the following parametric representations of the ``running'' of $\alpha(r)$ and $\lambda(r)$:
\begin{equation}
\label{alpha} \alpha(r) =\alpha_\infty(1-\exp(-r/{\bar r}))^{b/2},
\end{equation}
\begin{equation}
\label{lambda} \lambda(r)=\frac{\lambda_\infty}{{(1-\exp(-r/{\bar r}))^b}}.
\end{equation}
Here, ${\bar r}$ is a non-running distance scale parameter and $b$ is a constant.

\begin{figure}[ht]
\begin{center}
\input{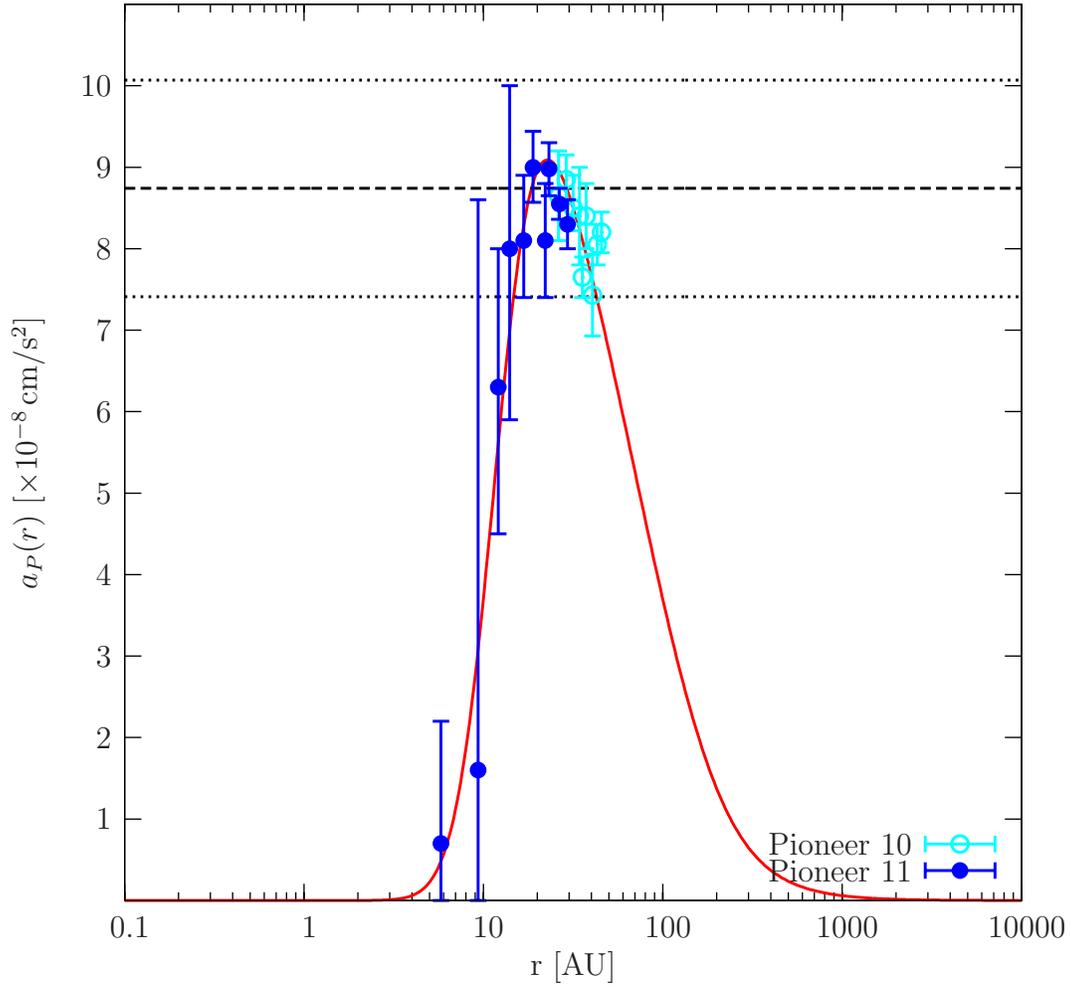}
\end{center}
\caption[Best fit to the Pioneer 10/11 anomalous acceleration data extracted from Figure 4 of \citeasnoun{Anderson3}
plotted against the position, $r$ in AU, on a logarithmic scale out to $r = 10,000 \au$.  ]{Best fit to the Pioneer 10/11 anomalous acceleration data extracted from Figure 4 of \citeasnoun{Anderson3}
plotted against the position, $r$ in AU, on a logarithmic scale out to $r = 10,000 \au$.  Pioneer 10 data is shown with
open cyan circles and Pioneer 11 data is shown with closed blue circles.  The Pioneer 10/11 anomalous acceleration,
$a_{P}$, is in units of $10^{-8}\ \mbox{cm/s}^{2}$ and the constant Pioneer anomaly result of \Eref{aP} is shown with
dotted black lines.}
\label{aulog}
\end{figure}

\begin{figure}[ht]
\begin{center}
\input{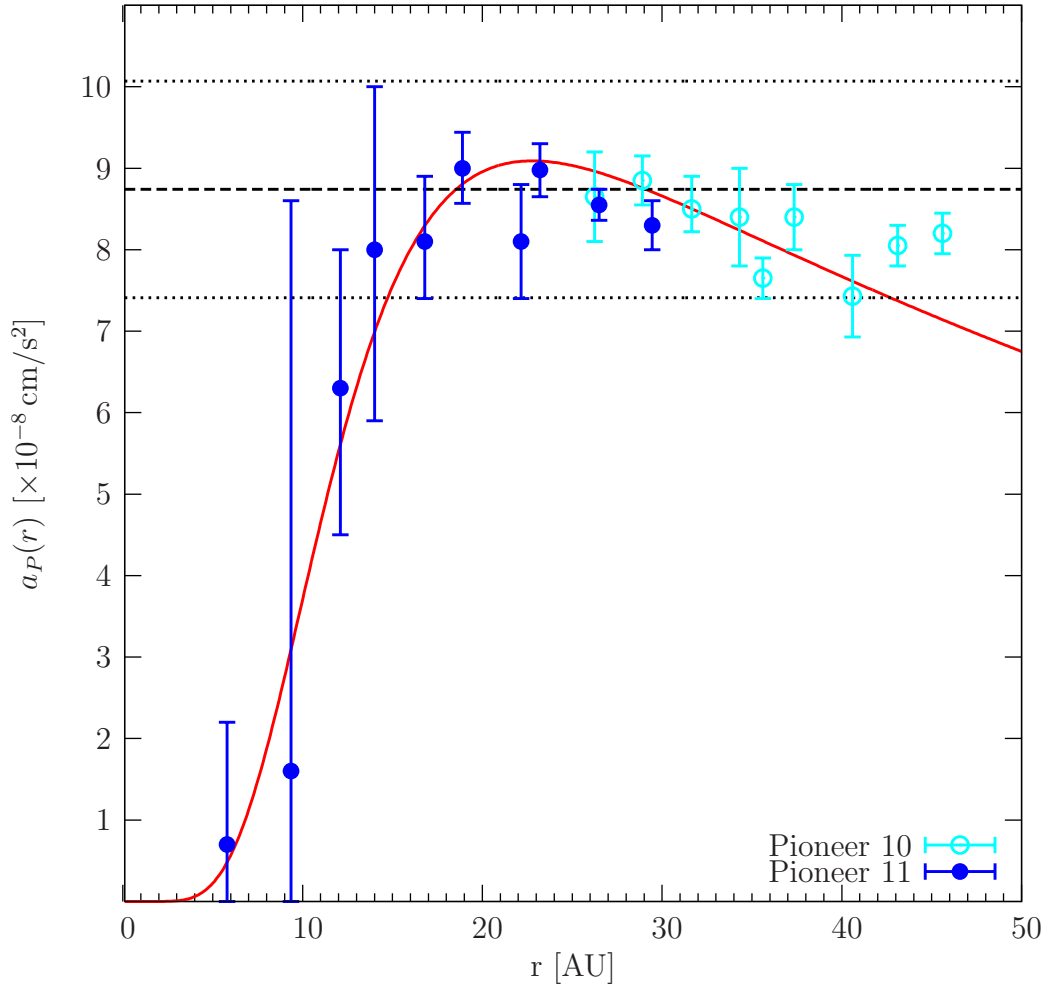}
\end{center}
\caption[Best fit to the Pioneer 10/11 anomalous acceleration data extracted from Figure 4 of \citeasnoun{Anderson3}
plotted against the position, $r$ in AU, on a linear scale out to $r = 50 \au$.  ]{ Best fit to the Pioneer 10/11 anomalous acceleration data extracted from Figure 4 of \citeasnoun{Anderson3}
plotted against the position, $r$ in AU, on a linear scale out to $r = 50 \au$.  Pioneer 10 data is shown with
open cyan circles and Pioneer 11 data is shown with closed blue circles.  The Pioneer 10/11 anomalous acceleration,
$a_{P}$, is in units of $10^{-8}\ \mbox{cm/s}^{2}$ and the constant Pioneer anomaly result of \Eref{aP} is
shown with dotted black lines.}
\label{aulin}
\end{figure}

In Figures \ref{aulog} and \ref{aulin}, we display a best fit to the acceleration data extracted from Figure 4 of \citeasnoun{Anderson3}
obtained using a nonlinear least-squares fitting routine including estimated errors from the Doppler shift
observations~\cite{Anderson2}.  The best fit parameters are:
\begin{eqnarray}
\nonumber \alpha_\infty &=& (1.00\pm0.02)\times 10^{-3},\\
\nonumber \lambda_\infty &=& 47\pm 1\au ,\\
\nonumber {\bar r} &=& 4.6\pm 0.2\au ,\\
\label{bestparameters} b &=& 4.0.
\end{eqnarray}
The small uncertainties in the best fit parameters are due to the remarkably low variance of residuals corresponding 
to a reduced $\chi^{2}$ per degree of freedom of 0.42 signalling a good fit.

Fifth force experimental bounds plotted for $\log_{10}\alpha$ versus $\log_{10}\lambda$ are shown in Fig. 1 of \citeasnoun{Reynaud} for fixed values of $\alpha$ and $\lambda$.
The updated 2003 observational data for the bounds obtained from the planetary ephemerides is extrapolated to $r = 10^{15}\,\mbox{m}=6,685 \au$~\cite{Fischbach2}. However, this extrapolation is based on using fixed universal
values for the parameters $\alpha$
and $\lambda$. Since known reliable data from the ephemerides of the outer planets ends with the data for Pluto at a
distance from the Sun, $r=39.52 \au=5.91\times 10^{12}\,m$, we could claim that for our range of values $47 \au <
\lambda(r) < \infty$, we predict $\alpha(r)$ and $\lambda(r)$ values consistent with the {\it un-extrapolated} fifth
force bounds.

A consequence of a variation of $G$ and $GM_\odot$ for the solar system is a modification of Kepler's third law:
\begin{equation}
\label{Kepler} a_{PL}^3=G(a_{PL})M_\odot\biggl(\frac{T_{PL}}{2\pi}\biggr)^2,
\end{equation}
where $T_{PL}$ is the planetary sidereal orbital period and $a_{PL}$ is the physically measured semi-major axis of the planetary orbit. For given
values of $a_{PL}$ and $T_{PL}$, \Eref{Kepler} can be used to determine $G(r)M_\odot$. The standard method is to use
astrometric data to define $GM_\odot$ for a constant value, 
\begin{equation} \label{gauss0}
G(r)M_\odot=G(a_{\oplus})M_\odot=\kappa^2,
\end{equation}
where $a_{\oplus}$ is the semi-major axis for Earth's orbit about the Sun, and $\kappa$ is the Gaussian gravitational constant given
by\footnote[1]{http://ssd.jpl.nasa.gov/astro\_constants.html}
\begin{equation} \label{gauss}
\kappa=0.01720209895 \au^{3/2}/{\rm day}.
\end{equation}
We obtain the standard semi-major axis value at 1 AU:
\begin{equation} \label{abar}
{\bar a}_{PL}^3 =G(a_\oplus)M_\odot\biggl(\frac{T_{PL}}{2\pi}\biggr)^2.
\end{equation}
For several planets such as Mercury, Venus, Mars and Jupiter there are planetary ranging data, spacecraft tracking data and
radiotechnical flyby observations available, and it is possible to measure $a_{PL}$ directly. For a distance varying $GM_\odot$ we
derive~\cite{Fischbach,Talmadge}:
\begin{equation}
\label{eta0}
\biggl(\frac{a_{PL}}{{\bar a}_{PL}}\biggr)=1+\eta_{PL} =\biggl[\frac{G(a_{PL})M_\odot}{\kappa^2}\biggr]^{1/3}.
\end{equation}
Here, it is assumed that $GM_\odot$ varies with distance such that $\eta_{PL}$ can be treated as a constant for the
orbit of a planet.  We may substitute the Gaussian gravitational constant of \Eref{gauss0} into \Eref{eta0} and obtain
\begin{equation}
\label{eta}
\eta_{PL} =\left[\frac{G(a_{PL})}{G(a_{\oplus})}\right]^{1/3}-1.
\end{equation}

\begin{figure}[ht]
\begin{center}
\input{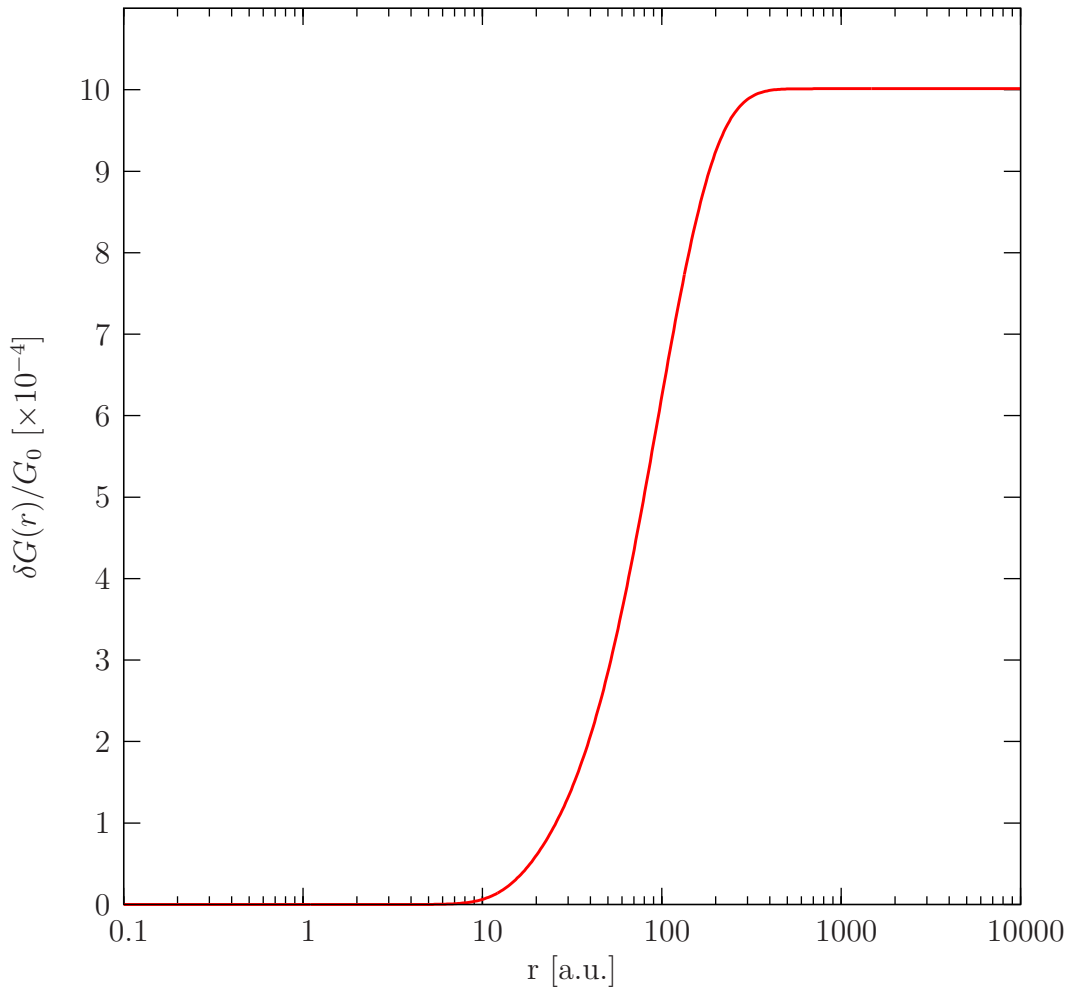}
\end{center}
\caption[Plot of $\delta G(r)/G_0$ arising from \Eref{deltaG} versus $r$ in AU predicted by the best fit
parameters of \Eref{bestparameters} to the Pioneer 10/11 anomalous acceleration data.]{ Plot of $\delta G(r)/G_0$ arising from \Eref{deltaG} versus $r$ in AU predicted by the best fit
parameters of \Eref{bestparameters} to the Pioneer 10/11 anomalous acceleration data.}
\label{runningG}
\end{figure}

In \Fref{runningG}, we display the variation of $\delta G/G_0$ arising from \Eref{deltaG} versus $r$ for
the parametric values of
$\alpha(r)$ and $\lambda(r)$ of \Eref{alpha} and \Eref{lambda}, respectively, using the best fit values for the parameters given in
\Eref{bestparameters}.
The behaviour of $G(r)/G_0$ is closely constrained to unity over the inner planets until beyond the orbit
of Saturn ($r \gtrsim 10 \au$) where the deviation in Newton's constant increases to an asymptotic value of
$G_{\infty}/G_{0} \rightarrow 1.001$ over a distance of hundreds of $\au$.  The transition to the renormalized Newtonian
constant is somewhat sudden as though the underlying physics was mimicking some source of dark matter. 
However, the fit is dictated by observations within the solar system and it is the nature of the renormalization group
flow to predict a continuous phase transition in the behaviour of the coupling constant in the vicinity of the infrared
fixed point.  It is this underlying physics which we have applied phenomenologically to solve the problem of galaxy
rotation curves without dark matter~\cite{Brownstein} and galaxy cluster masses without dark matter~\cite{Brownstein2}.  

\begin{table}[ht]
\caption[Theoretical predictions of the values of $\eta_{PL}$ of \Eref{eta} 
compared to the observational limits taken from \citeasnoun{Talmadge} and the best-fit
theoretical predictions for the Pioneer Anomaly, $a_{P}$, for the nine planets at a distance $r$ from the
Sun.]{Theoretical predictions of the values of $\eta_{PL}$ of \Eref{eta} 
compared to the observational limits taken from \citeasnoun{Talmadge} and the best-fit
theoretical predictions for the Pioneer Anomaly, $a_{P}$, for the nine planets at a distance $r$ from the
Sun.  No observational limits were computed beyond Saturn in \citeasnoun{Talmadge} due to uncertainty in optical data. 
Beyond the outer planets, the theoretical
predictions for $\eta(r)$ approaches the asymptotic value $\eta_{\infty} = 3.34 \times 10^{-4}$.}
\label{predictions}
\begin{tabular}{@{}l|cccc}
\br
Planet & r &Theoretical Prediction & Observational Limit & $a_{P}$ \\
&(AU)&$\eta_{PL}\ (10^{-10})$ &$\eta_{PL}\ (10^{-10})$ & $(10^{-8}\ \mbox{cm/s}^{2})$\\
\mr
Mercury & \lineup\00.38 & $-6.55 \times 10^{-5}$ & $+40\pm50$ & $1.41 \times 10^{-10}$\\
Venus & \lineup\00.72 & $-6.44 \times 10^{-5}$ & $-55\pm35$ & $5.82 \times 10^{-8}$\\
Earth & \lineup\01.00 & \lineup\m$0.00\times 10^{0\lineup\m} $ & $\lineup\m0$ & $1.16 \times 10^{-6}$\\
Mars & \lineup\01.52 & $\lineup\m4.93 \times 10^{-3}$ & $-0.9\pm2.1$ & $4.42 \times 10^{-5}$ \\
Jupiter & \lineup\05.20 & $\lineup\m4.19 \times 10^{2\lineup\m}$ & $+200\pm400$ & $2.76 \times 10^{-1}$\\
Saturn & \lineup\09.54 & $\lineup\m1.67 \times 10^{4\lineup\m}$ & \ldots & $3.27 \times 10^{0\lineup\m}$ \\ 
Uranus & 19.22 & $\lineup\m1.84 \times 10^{5\lineup\m}$ & \ldots & $8.86 \times 10^{0\lineup\m}$ \\
Neptune & 30.06 & $\lineup\m4.39 \times 10^{5\lineup\m}$ & \ldots & $8.65 \times 10^{0\lineup\m}$ \\
Pluto & 39.52 & $\lineup\m6.77 \times 10^{5\lineup\m}$ & \ldots & $7.72 \times 10^{0\lineup\m}$ \\
\br
\end{tabular}
\end{table}
For the nine planets, we obtain the values of $\eta_{PL}$ shown in \Tref{predictions}. We see that we are able to
obtain agreement well within the bounds of possible variation of $GM_\odot$ consistent with the data~\cite{Fischbach,Talmadge}
for Mercury, Venus, Mars and Jupiter.  No observational limit on $\eta_{PL}$ for Saturn or the outer planets has yet
been established; but this is precisely where the deviation $\delta G(r)/G_0$ leads to a sizable contribution in the
theoretical prediction for $\eta_{PL}$.  The reason for the uncertainty beyond the orbit of Saturn and the lack of
observational limits on $\eta_{PL}$ is that the ephemerides for the outer planets is based on optical measurements. 
Even in the context of Newton's theory, the extrapolation of Kepler's third law of \Eref{Kepler} using the Gaussian
gravitational constant of \Eref{gauss} which fits the inner planets misestimates the semi-major axis, $a_{PL}$, or the
orbital period, $T_{PL}$, of the outer planets resulting from Newtonian perturbations due to Jupiter and the gas giants
and their satellites, the Kuiper belt and hundreds of asteroids.  The latest version of the planetary part of the numerical ephemerides is a
numerical integration of the post-Newtonian metric.  It attempts to account for these perturbations from Kepler's law
beyond Saturn by a least squares adjustment to all the available observations including the CCD optical astrometric
observations of the outer planets.  These values (without uncertainty) are
available from the Solar System Dynamics Group (SSD) of the Jet Propulsion Laboratory (JPL) through the Horizon's
ephermeris DE410 online\footnote[2]{http://ssd.jpl.nasa.gov/horizons.html}. The Russian Academy of Sciences has also
placed their latest values known as
EPM2004 online\footnote[3]{ftp://quasar.ipa.nw.ru/incoming/EPM2004}.  Because the perturbations change daily due to the
motion within the
solar system, the planetary ephemerides quoted values for $a_{PL}$ and $T_{PL}$ change daily.  In order to compute
deviations from Kepler's third law for the outer planets, we
have listed today's best known values in \Tref{ephemerides}.  The uncertainty in the
EPM2004 deduced values for the semi-major axes of the planets, $\Delta a_{PL}$, have been studied in \citeasnoun{Pitjeva}
and the quoted values are listed in \Tref{ephemerides}.  \citeasnoun{Pitjeva} warns that the real errors
may be larger by an order of magnitude.  The uncertainty in the periods for the outer planets are not quoted in either
EPM2004 or DE410, and so we have assumed small uncertainties based on the precision provided by the JPL Horizon's
online ephemeris.  We may calculate the uncertainty, $\Delta \eta_{PL}$,
by propagating the errors $\Delta a_{PL}$ and $\Delta T_{PL}$ according to Equations \eref{abar} and \eref{eta0},
neglecting any uncertainty in the Gaussian graviational constant of \Eref{gauss}:
\begin{equation}
\label{Deta}
\Delta \eta_{PL} = \sqrt{\left(\frac{\Delta a_{PL}}{\bar a_{PL}}\right)^{2}+\left(\frac{2}{3}\frac{a_{PL}}{\bar a_{PL}}\frac{\Delta
T_{PL}}{T_{PL}}\right)^{2}}.
\end{equation}
\begin{table}[b]
\caption[Mean ephemerides orbital parameters for the semi-major axes (J2000), orbital eccentricities and the sidereal
orbital periods (JPL 
Horizon's online ephemeris).]{Mean ephemerides orbital parameters for the semi-major axes (J2000), orbital eccentricities and the sidereal
orbital periods (JPL 
Horizon's online ephemeris).  The errors in the semi-major axes are deduced from Table 4 of \citeasnoun{Pitjeva} with
1 AU = $(149597870696.0 \pm 0.1)$ m.  The computation of $\Delta
\eta_{PL}$ is based on the propagation of the uncertainties according to \Eref{Deta}.}
\label{ephemerides}
\begin{tabular}{@{}l|llllll}
\br
Planet & $a_{PL}$ & $\Delta a_{PL}$ & $e_{PL}$ &$T_{PL}$ & $\Delta T_{PL}$ & $\Delta \eta_{PL}$  \\
&(AU)&(AU)&&(days)&(days)&$(10^{-10})$\\
\mr
Mercury & \lineup\00.38709893 & $7.02 \times 10^{-13}$ & 0.206 & \lineup\0\0\087.968435 & $5.0 \times 10^{-7}$ & $3.79\times
10^{1}$ \\
Venus & \lineup\00.72333199 & $2.20 \times 10^{-12}$ & 0.007 & \lineup\0\0224.695434 & $5.0 \times 10^{-7}$ & $1.48\times
10^{1}$ \\
Earth & \lineup\01.00000011 & $9.76 \times 10^{-13}$ & 0.017 & \lineup\0\0365.256363051 & $5.0 \times 10^{-10}$ & $1.33\times 10^{-2}$ \\
Mars & \lineup\01.52366231 & $4.39 \times 10^{-12}$ & 0.093 & \lineup\0\0686.980 & $5.0 \times 10^{-4}$ & $4.85\times
10^{3}$ \\
Jupiter & \lineup\05.20336301 & $4.27 \times 10^{-9}$ & 0.048 & \lineup\04330.595 & $5.0 \times 10^{-4}$ & $7.68\times
10^{2}$ \\
Saturn & \lineup\09.53707032 & $2.82 \times 10^{-8}$ & 0.056 & 10746.94 & $5.0 \times 10^{-2}$ & $3.10\times 10^{4}$ \\ 
Uranus & 19.19126393 & $2.57 \times 10^{-7}$ & 0.047 & 30685.4 & $1.0 \times 10^{0}$& $2.17\times 10^{5}$ \\
Neptune & 30.06896348 & $3.20 \times 10^{-6}$ & 0.009 & 60189. & $5.0 \times 10^{0}$ & $5.54\times 10^{5}$ \\
Pluto & 39.48168677 & $2.32 \times 10^{-5}$ & 0.250 & 90465. & $1.0 \times 10^{1}$ &$7.37\times 10^{5}$ \\
\br
\end{tabular}
\end{table}

Although according to \Tref{predictions} we are consistent with the observational limits of  $\eta_{PL}$ for the%
inner planets to Jupiter, the computation of \citeasnoun{Talmadge} attempted
to set model-independent constraints on the possible modifications of Newtonian gravity. 
The procedure was to run the planetary ephemerides numerical integration with the addition of $\eta_{PL}$ as free
parameters.  Because there was one additional parameter for each planet, they were only able to find observational limits
for the inner planets including Jupiter.  In order to compute the observational limit for $\eta_{PL}$ for the outer planets, it would be necessary to
compute the planetary ephemerides using the modified acceleration law of Equations \eref{accelerationGrun} and \eref{runningNewton}. 
Although this is beyond the scope of the current investigation, we may approximate here the observational limit of
$\eta_{PL}$ for the outer planets as the uncertainty $\Delta \eta_{PL}$ from \Eref{Deta}, for the perturbations
of \Fref{runningG}, $\delta G(r)/G_0$, are small compared to the Newtonian perturbations
acting on the outer planets. The results for $\Delta \eta_{PL}$ due to the uncertainty in the planetary ephemerides are presented in \Tref{ephemerides} for the nine planets and exceed 
the predictions, $\eta_{PL}$, of \Tref{predictions}.

The relativistic equation of motion for a test particle in our gravitational theory may be solved
perturbatively in a weak field approximation for 
the anomalous perihelion advance of a planetary orbit:
\begin{equation}
\label{perihelionAdvance}\Delta\omega_{PL}=\frac{6\pi G_{0}M_\odot}{c^2a_{PL}(1-e_{PL}^2)}(1-\alpha_{PL}),
\end{equation}
where we have assumed as with Kepler's third law that $GM_\odot$ and $\alpha$ vary with distance such that they can
be
treated as constants for the orbit of a planet, where we have made use of the approximation $G(r) \approx
G_{0}$~\cite{Moffat2}, which is the case from the fit to the Pioneer 10/11 anomalous acceleration data.
We may rewrite
\Eref{perihelionAdvance} as the perihelion advance in arcseconds per century:
\begin{equation}
\label{perihelionAdvancedot}{\dot\omega_{PL}}=\frac{\Delta\omega_{PL}}{2\pi T_{PL}} = \frac{3 G_{0}M_\odot}{c^2a_{PL}(1-e_{PL}^2) T_{PL}}(1-\alpha_{PL}),
\end{equation}
where $T_{PL}$ is the planetary orbital period, and $e_{PL}$ is the planetary orbital eccentricity.  
We may separate
\Eref{perihelionAdvancedot} into the usual Einstein anomalous perihelion advance, and a prediction of the correction
to the anomalous perihelion advance:
 \begin{equation}
\label{perihelionAdvance.split}{\dot \omega_{PL}}={\dot \omega_{0}} +{\dot \omega_{1}},
\end{equation} 
\begin{table}[ht]
\caption[The values of the running parameters, $\alpha(r)$ of \Eref{alpha} and $\lambda(r)$ of \Eref{lambda} and the deviation in the dimensionless gravitational constant, $\delta G(r)/G_{0}$ of \Eref{runningNewton}, calculated for each planet.]
{The values of the running parameters, $\alpha(r)$ of \Eref{alpha} and $\lambda(r)$ of \Eref{lambda} and the deviation in the dimensionless gravitational constant, $\delta G(r)/G_{0}$ of \Eref{runningNewton}, calculated for each planet.  
Included on the right of the table is the theoretical (Einstein) perihelion advance of \Eref{perihelionAdvance.Einstein},
and the predicted retrograde of
\Eref{perihelionAdvance.Retrograde} for the planets, and the limits set by the ephemeris~\cite{Pitjeva2}.\label{perihelia}
\label{table.perihelion}}
\begin{tabular}{@{}l|ccc|ccc}
\br
&&&&\multicolumn{3}{|c}{ ${\dot \omega}\ (^{\prime\prime}/$century)}  \\
Planet & $\alpha_{PL}$ & $\lambda_{PL}\ (\au)$ & $\delta G_{PL}/G_{0}$ & Einstein & Retrograde & Ephemeris  \\
\mr
Mercury & $6.51 \times 10^{-6}$ & $1.11 \times 10^6$ & $3.98 \times 10^{-19}$ & $42.99$& $-2.80 \times\ 10^{-4}$
&$-0.0336\pm0.0050$ \\
Venus & $2.12 \times 10^{-5}$ & $1.05 \times 10^5$ & $5.04 \times 10^{-16}$ & $8.63$ & $-1.83 \times\ 10^{-4}$  & \ldots\\
Earth & $3.82 \times 10^{-5}$ & $3.23 \times 10^4$ & $1.84 \times 10^{-14}$ & $3.84$ & $-1.47 \times\ 10^{-4}$  &$-0.0002\pm 0.0004$\\
Mars & $7.95 \times 10^{-5}$ & $7.44 \times 10^3$ & $1.67 \times 10^{-12}$ & $1.35$ & $-1.07 \times\ 10^{-4}$  &$0.0001\pm 0.0005$\\
Jupiter & $4.59 \times 10^{-4}$ & $2.23 \times 10^2$ & $1.23 \times 10^{-7}$ & $0.0624$ & $-2.86 \times\
10^{-5}$  &$0.0062\pm 0.036$ \\
Saturn & $7.64 \times 10^{-4}$ & $8.05 \times 10^1$ & $4.96 \times 10^{-6}$ & $0.0137$ & $-1.05 \times\
10^{-5}$  &$-0.92\pm 2.9$\\
Uranus & $9.69 \times 10^{-4}$ & $5.00 \times 10^1$ & $5.55 \times 10^{-5}$  & $0.00239$ & $-2.31 \times\
10^{-6}$  &$0.57\pm 13.0$\\
Neptune & $9.97 \times 10^{-4}$ & $4.73 \times 10^1$ & $1.34 \times 10^{-4}$ & $0.00078$ & $-7.73 \times\
10^{-7}$  &
\ldots\\
Pluto & $1.00 \times 10^{-3}$ & $4.70 \times 10^1$ & $2.05 \times 10^{-4}$ & $0.00042$ & $-4.18 \times\ 10^{-7}$  & \ldots\\
\br
\end{tabular}
\end{table}
where \begin{equation}
\label{perihelionAdvance.Einstein}{\dot \omega_{0}} = \frac{3 G_{0}M_\odot}{c^2a_{PL}(1-e_{PL}^2)T_{PL}}
\end{equation} is the Einstein anomalous perihelion advance, and \begin{equation}
\label{perihelionAdvance.Retrograde}{\dot \omega_{1}} = -\alpha_{PL} {\dot \omega_{0}}
\end{equation} is the predicted retrograde (note the {\it minus-sign} in \Eref{perihelionAdvance.Retrograde}) to the
anomalous perihelion advance of \Eref{perihelionAdvance.Einstein}. The measured perihelion precession is
best known for the inner planets  (for Mercury the precession obtained from ranging data is known to $0.5\%$~\cite{Will}).  For each of the planets in the solar system, we find that
$\alpha_{PL} << 1$, so that our fit to the Pioneer anomalous acceleration is in agreement with the relativistic
precession data.  The results for the Einstein perihelion advance, and our predicted retrograde for each planet, and the
observational limits set by the recent ultra-high precision ephemeris are listed in \Tref{table.perihelion}.

The validity of the bounds on a possible fifth force obtained from the ephemerides of the outer planets Uranus, Neptune
and Pluto are critical in the exclusion of a parameter space for our fits to the Pioneer anomaly acceleration. Beyond
the outer planets, the theoretical prediction for $\eta(r)$ approaches an asymptotic value:
\begin{equation} \label{etalim}
\eta_{\infty} \equiv \lim_{r \to \infty} \eta(r)= 3.34 \times 10^{-4}.
\end{equation}
We
see that the variations (``running'') of $\alpha(r)$ and $\lambda(r)$ with distance play an important role in
interpreting the data for the fifth force bounds. This is in contrast to the standard non-modified Yukawa correction to
the Newtonian
force law with fixed universal values of $\alpha$ and $\lambda$ and for the range of values $0 < \lambda < \infty$, for which the
equivalence principle and lunar laser ranging and radar ranging data to planetary probes exclude the possibility of a
gravitational and fifth force explanation for the Pioneer anomaly.

\section{Conclusions}

A modified gravity theory based an a $D=4$ pseudo-Riemannian metric, a spin-$1$ vector field and a corresponding skew
symmetric field $B_{\mu\nu}$ and dynamical scalar fields $G$, $\omega$ and $\mu$, yields a static spherically symmetric
gravitational field with an additional Yukawa potential and with effective varying coupling strength $\alpha(r)$ and
distance range $\lambda(r)$. This modified acceleration law leads to remarkably good fits to a large number of
galaxies~\cite{Brownstein} and galaxy clusters~\cite{Brownstein2} without non-baryonic dark matter. The previously published gravitational theories
MSTG~\cite{Moffat} and STVG~\cite{Moffat2} yielded the same modified weak gravitational field acceleration law and,
therefore, the same successful fits to galaxy and cluster data. The MSTG and STVG gravity theories can both be
identified generically as metric-skew-tensor gravity theories, for they both describe gravity as a metric theory with an
additional degree of freedom associated with a skew field coupling to matter. For MSTG, this is a third-rank skew field
$F_{\mu\nu\lambda}$, while for STVG the skew field is a second-rank tensor $B_{\mu\nu}$.

An action $S_S$ for the scalar fields $G(x)$, $\omega(x)$ and $\mu(x)=1/\lambda(x)$ and the field equations resulting from a
variation of the action, $\delta S_S=0$, can be incorporated into the MSTG and STVG theories. The dynamical solutions
for the scalar fields give an {\it effective} description of the running of the constants in an RG flow quantum gravity
scenario~\cite{Reuter,Moffat}, in which strong infrared renormalization effects and increasing large scale spatial
values of $G$ and $\omega$ lead, together with the modified acceleration law, to a satisfactory description of galaxy
rotation curves and cluster dynamics without non-baryonic dark matter, as well as a solution to the Pioneer 10/11
anomalous acceleration data.

We have demonstrated that the STVG theory can explain the Pioneer anomalous acceleration data and still be consistent
with the accurate equivalence principle, lunar laser ranging and satellite data for the inner solar system as well as
the outer solar system planets including Pluto at a distance of $r= 39.52 \au =5.91\times 10^{12}$ meters. The ephemerides
for the outer planets are not as well know as the ones for the inner planets due to their large distances from the Sun.
The orbital data for Pluto only correspond to the planet having gone round 1/3 of its orbit. It is important that the
distance range parameter lies in the region $47 \au <
\lambda(r) < \infty$ for the best fit to the Pioneer acceleration data, for the range in the modified Yukawa correction
to Newtonian gravity lies in a distance range beyond Pluto. Further investigation of fifth force bounds obtained by an
analysis of the planetary data for the outer planets, based on the modified gravity theory is required.  We are
predicting that measurements of a fifth force in the solar system will become measurable at distances $r \gtrsim  10
\au$ from the Sun where as shown in \Fref{runningG}, $\delta G(r)/G_0$ (and $\eta_{PL}$) become potentially
measureable.

Perhaps, a future deep space probe can produce data that can check the predictions obtained for the Pioneer anomaly from
our modified gravity theory. Or perhaps utilizing minor planets may clarify whether the Pioneer
anomaly is caused by the gravitational field in the outer solar system~\cite{Page}.  An
analysis of anomalous acceleration data obtained from earlier Doppler shift data
retrieval will clarify in better detail the apparent onset of the anomalous acceleration beyond the position of Saturn's
orbit. The planned LATOR mission will achieve significantly improved measurements of possible new degrees of freedom
such as dynamical scalar and vector fields in the solar system, which could test the proposed MSTG and STVG
theories~\cite{Turyshev2}.

\ack 
This work was supported by the Natural Sciences and Engineering Research Council of Canada. We thank Eric
Adelberger, Ephraim Fischbach, Martin Green, Jo\~ao Magueijo and Michael Nieto for helpful discussions.
\addcontentsline{toc}{section}{Acknowledgments}


\addcontentsline{toc}{section}{References}
\Bibliography{10}
\bibitem{Anderson} Anderson\,J.\,D., Laing\,P.\,A.,  Lau\,E.\,L., Liu\,A.\,S., Nieto\,M.\,M. and Turyshev\,S.\,G. 1998 \PRL {\bf 81} 2858--61 (\preprint{gr-qc/9808081}) 
\bibitem{Anderson2} Anderson\,J.\,D., Laing\,P.\,A.,  Lau\,E.\,L., Liu\,A.\,S., Nieto\,M.\,M. and Turyshev\,S.\,G. 2002 \PR D {\bf 65} 082004 (\preprint{gr-qc/0104064}) 
\bibitem{Turyshev} Turyshev\,S.\,G., Nieto M.\,M.\, and Anderson\,J.\,D. 2005 (\preprint{gr-qc/0510081})  
\bibitem{Moffat} Moffat\,J.\,W. 2005 \JCAP JCAP05(2005)003 (\preprint{astro-ph/0412195}) 
\bibitem{Moffat2} Moffat\,J.\,W. 2006 \JCAP JCAP03(2006)004 (\preprint{gr-qc/0506021}) 
\bibitem{Brownstein} Brownstein\,J.\,R. and Moffat\,J.\,W. 2006  \ApJ {\bf 635} 721--741  (\preprint{astro-ph/0506370}) 
\bibitem{Brownstein2} Brownstein\,J.\,R. and Moffat\,J.\,W.  2006 \MNRAS {\bf 367} 527--540 \\
(\preprint{astro-ph/0507222}) 
\bibitem{Reuter} Reuter\,M. and Weyer\,H. 2004 \JCAP JCAP12(2004)001 \\
(\preprint{hep-th/0410119}) 
\bibitem{Fischbach} Fischbach\,E. and Talmadge\,C.\,L. 1999 {\it The Search for Non-Newtonian Gravity} (Heidelberg -- New York: Springer)  
\bibitem{Talmadge} Talmadge\,C., Berthias\,J.\,P., Hellings\,R.\,W. and Standish\,E.\,M. 1988 \PRL {\bf 61} 1159 
\bibitem{Adelberger} Stubbs\,C.\,W., Adelberger\,E.\,G., Raab\,F.\,J., Gundlach\,J.\,H., Heckel\,B.\,R., McMurry\,K.\,D., Swanson\,H.\,E. and Watanabe\,R. 1987 \PRL {\bf 58} 1070--3 
\bibitem{Adelberger1} Adelberger\,E.\,G., Stubbs\,C.\,W., Rogers\,W.\,F., Raab\,F.\,J., Heckel\,B.\,R., Gundlach\,J.\,H., Swanson\,H.\,E. and Watanabe\,R. 1987  \PRL {\bf 59} 849--52 
\bibitem{Adelberger2} Adelberger\,E\,G, Stubbs\,C\,W, Heckel\,B\,R, Su\,Y, Swanson\,H\,E, Smith\,G and Gundlach\,J\,H 1990 \PR D {\bf 42} 3267--92 
\bibitem{Adelberger3} Adelberger\,E.\,G., Heckel and Nelson\,A.\,E. 2003 {\it Ann.\,Rev.\,Nucl.\,Part.\,Sci.} {\bf 53} 77--121\\
(\preprint{hep-ph/0307284}) 
\bibitem{Reynaud} Reynaud\,S. and Jaekel\,M.\,M. 2005 {\it Int.\,J.\,Mod.\,Phys.} {\bf A20} 2294--2303  (\preprint{gr-qc/0501038}) 
\bibitem{Anderson3} Nieto\,M.\,M. and Anderson\,J.\,D. 2005 \CQG {\bf 22} 5343-5354 \\
(\preprint{gr-qc/0507052}) 
\bibitem{Fischbach2} Fischbach\,E.  2005 private communication 
\bibitem{Pitjeva} Pitjeva\,E.\,V. 2005 \SSR {\bf 39} 202--213 
\bibitem{Pitjeva2} Pitjeva\,E.\,V. 2005 {\it Astron.\,Lett.} {\bf 31} 340--349 
\bibitem{Will} Will\,C.\,W. 2006 {\it Living Rev.\,Relativity} {\bf 9} (\preprint{gr-qc/0510072}) 
\bibitem{Page} Page\,G.\,L., Dixon\,D.\,S., and Wallin\,J.\,F. 2006  \ApJ {\bf 642} 606--614 (\preprint{astro-ph/0504367})
\bibitem{Turyshev2} Turyshev\,S.\,G., Shao\,M. and Nordtvedt\,K.\,L. 2004 {\it Int.\,J.\,Mod.\,Phys.\,D} {\bf 13} 2035--64\\
(\preprint{gr-qc/0410044})
\endbib
\end{document}